\documentclass{article}
\usepackage{amsmath}
\usepackage{graphicx}

\begin{document}

\begin{center}

{\Large \textbf{ Cavity equations for a positive or negative  refraction index  
material with electric and magnetic non-linearities
}}

\vspace{0.6cm}

 Daniel A. M\'{a}rtin and Miguel Hoyuelos
  
 \emph{Departamento de F\'{\i}sica, Facultad de Ciencias Exactas y Naturales,\\ Universidad Nacional de Mar del Plata \\ and Instituto de Investigaciones F\'{\i}sicas de Mar del Plata\\ (Consejo Nacional de Investigaciones Cient\'{\i}ficas y T\'{e}cnicas),\\ Funes 3350, 7600 Mar del Plata, Argentina}

\end{center}
\vspace{0.2cm}
\begin{abstract}
We study evolution equations for electric and magnetic
field amplitudes in a ring cavity with plane mirrors.  The
cavity is filled with a positive or negative refraction
index material with third order effective electric and
magnetic non-linearities.  Two coupled non-linear equations
for the electric and magnetic amplitudes are obtained.  We
prove that the description can be reduced to one Lugiato
Lefever equation with generalized coefficients.  A
stability analysis of the homogeneous solution,
complemented with numerical integration, shows that any
combination of the parameters should correspond to one of
three characteristic behaviors.
\end{abstract}



\section{Introduction}
\label{intr}

During the last decade, composite material developments allowed the
experimental realization of materials with negative refraction index
\cite{smith,shelby}, theoretically predicted by Veselago
\cite{veselago}.  These materials have simultaneously a negative
dielectric permittivity and negative magnetic permeability, a
property that is not found in natural materials. Negative refraction
index materials (NRM) have many new and interesting properties, such
as a refracted wave on the same side as the incoming wave respect to
the surface normal and Poynting vector in the direction opposite to
wave-vector. A number of applications has been proposed based on
their new properties, such as perfect lenses, phase compensators, and
electrically small antennas \cite{jaksic,rama,grbic,fang}.

Most studies on NRM consider linear relations between electric field
and polarization, and magnetic field and magnetization. The study of
nonlinear effects acquired an increasing interest during the last
years.  It is known that the propagation of an electromagnetic wave
in a Kerr nonlinear material, with positive refraction index, can be
described by a nonlinear order parameter equation of the Schr\"{o}dinger
type \cite{moloney}, and that the same kind of equation can be
extended to an NRM \cite{kockaert,kockaert2,wen,scalora,lazarides}.
Nevertheless, it has been shown \cite{zharov,lapine} that a composite
metamaterial with negative refraction index can develop a non-linear
macroscopic magnetic response.  This means that, although the host
medium has a negligible magnetic non-linearity, the periodic
inclusions of the metamaterial produce an effective magnetic
non-linear response when the wave-length is much larger than the
periodicity of the inclusions. In the rest of the paper, when we
speak about magnetic non-linearity we refer to this macroscopic
magnetic response present only in metamaterials.  Electric and
magnetic non-linearities in composite materials have been also
analyzed in, for example, \cite{brien,shadrivov,shadrivov2,lapine2}

In this paper, we are interested in the analysis of the equations
that describe the electric and magnetic fields in a ring cavity with
plane mirrors containing a material with negative or positive
refraction index and with electric \emph{and} magnetic
non-linearities. The aim of the work is to obtain a simple
mathematical description of this system, useful to identify relevant
parameters and to analyze typical behaviours.

The paper is organized as follows.  In Sect.\ \ref{equ} we present
the equations for the evolution of the electric and magnetic field
amplitudes in a ring cavity (the derivation, starting from two
coupled non-linear Schr\"{o}dinger equations, is in the appendix), and
show that the description can be reduced to one Lugiato Lefever
equation \cite{lugiato} with generalized parameters. In Sect.\
\ref{dissipation}, we analyze the effects of dissipation.  In Sect.\
\ref{staban} we use a linear stability analysis to identify three
typical situations that arise depending on the signs of the three
parameters of the equation. Numerical integration supports and
completes the previous analysis. In Sect.\ \ref{concl} we present our
conclusions.

\section{Equations in the cavity}
\label{equ}

The equations that describe the behavior of the electric and magnetic
fields in a plane perpendicular to light propagation are of the type
of the Lugiato-Lefever (LL) equation. The LL equation is a simple
mean field model that has been useful for the analysis of pattern
formation in a cavity with flat mirrors containing a Kerr medium and
driven by a coherent plane-wave field, see also
\cite{firth,hoyuelos}.

We will first analyze the problem of free propagation (without
mirrors) of an electromagnetic wave in the material and afterwards
will use the resulting equations to derive the behavior in the
cavity.

We will consider a linearly polarized driving field with frequency
$\omega_0$. Let us suppose that the electric field is in the $x$
direction and the magnetic field is in the $y$ direction. The
starting point are the Maxwell's equations and the constitutive
relations for the electric displacement, $D = \epsilon_0 E + P$, and
the magnetic induction, $B=\mu_0 H + \mu_0 M$. (There is an
interesting alternative approach, described in \cite{agranovich}, in
which the fields $E$, $D$ and $B$ are used, with
$D=\tilde{\varepsilon} E$ and $B=H$, where $\tilde{\varepsilon}$ is a
generalized dielectric constant.)

Considering an isotropic metamaterial with third order
non-linearities, the non-linear relation between polarization $P$ (in
the $x$ direction) and electric field $E$ is
\begin{equation}
P(t) = \epsilon_0 \int_{-\infty}^\infty \chi^{(1)}_e(t-\tau)\,
E(\tau)\, d\tau + \epsilon_0 \int_{-\infty}^\infty
\chi^{(3)}_e(t-\tau_1,t-\tau_2,t-\tau_3)\,
E(\tau_1)E(\tau_2)E(\tau_3)\, d\tau_1 d\tau_2 d\tau_3, \label{pol}
\end{equation}
where $\chi^{(1)}_e$ and $\chi^{(3)}_e$ are the linear and nonlinear
electric susceptibilities \cite{moloney}.  The same geometric
arguments used in (\ref{pol}) can be applied to the relation between
magnetization $M$ and magnetic field $H$:
\begin{equation}
M(t) = \int_{-\infty}^\infty \chi^{(1)}_m(t-\tau)\, H(\tau)\, d\tau +
\int_{-\infty}^\infty \chi^{(3)}_m(t-\tau_1,t-\tau_2,t-\tau_3)\,
H(\tau_1)H(\tau_2)H(\tau_3)\, d\tau_1 d\tau_2 d\tau_3, \label{magn}
\end{equation}
with $\chi^{(1)}_m$ and $\chi^{(3)}_m$ being the linear and nonlinear
magnetic susceptibilities.  Let us remark that $\chi^{(1)}_e$,
$\chi^{(3)}_e$, $\chi^{(1)}_m$ and $\chi^{(3)}_m$ describe
macroscopic effective response of the metamaterial valid for a
driving field with a wave-length much larger than the periodicity of
the inclusions.  We are assuming that the field intensities are small
enough in order to neglect higher order terms.  The effective
macroscopic magnetic non-linearity can become relevant in
metamaterials, as shown in \cite{zharov}, in contrast with what
happens in conventional optics, where magnetic non-linearities are
negligible.

The classical multiple-scales perturbation technique will be applied,
in which it is assumed that the light is quasimonochromatic and can
be represented by a plane wave, with frequency $\omega_0$ and
wave-number $k_0$, propagating along the $z$ axis and modulated by a
slowly varying envelope. The envelope depends on space
$\mathbf{R}=(X,Y,Z)$ and time $T$ variables that have characteristic
scales much greater than the scales given by $1/k_0$ and
$1/\omega_0$. The fields $E$ and $H$ can be written in the following
way
\begin{eqnarray}
E &=& \mathcal{E}(\mathbf{R},T) e^{i(k_0 z - \omega_0 t)} + c.c. \nonumber \\
H &=& \mathcal{H}(\mathbf{R},T) e^{i(k_0 z - \omega_0 t)} + c.c.
\label{ansatz}
\end{eqnarray}
where $\mathcal{E}$ and $\mathcal{H}$ are the slowly varying
amplitudes. Similar relations hold for $P$ and $M$.

Let us define the wave-number $k(\omega) = \omega n(\omega)/c$; the
refraction index is
$n(\omega)=\pm\sqrt{\epsilon_r(\omega)\,\mu_r(\omega)}$ (it takes the
negative sign when both, $\epsilon_r$ and $\mu_r$ are negative
\cite{veselago}), where $\epsilon_r(\omega) = 1 +
\chi^{(1)}_e(\omega)$ and $\mu_r(\omega) = 1 + \chi^{(1)}_m(\omega)$
are the relative permittivity and the relative permeability
respectively. In particular, $k_0 = \omega_0 n/c$, with
$n=n(\omega_0)$, and we call $k'$ and $k''$ the derivatives of
$k(\omega)$ evaluated in $\omega=\omega_0$.

The details of the application of the multiple-scales technique in
our case are essentially the same to the ones described in Ref.\
\cite{moloney}, Sect. 2k, for a positive refraction index material
with only electric non-linearity. After this process, we arrive to
the following coupled non linear Schr\"{o}dinger equations for the
envelopes of the electric and magnetic fields:
\begin{eqnarray}
\frac{\partial \mathcal{E}}{\partial \xi} &=& - \frac{i k''}{2}
\frac{\partial^{2} \mathcal{E}}{\partial t^2} + \frac{i}{2k_0}
\nabla_\perp^{2}\mathcal{E} + \frac{i 3 k_0}{2}
\left(\frac{\chi^{(3)}_e}{\epsilon_r}|\mathcal{E}|^{2} +
\frac{\chi^{(3)}_m}{\mu_r}|\mathcal{H}|^{2}\right)\mathcal{E}
\label{eampl}\\
\frac{\partial \mathcal{H}}{\partial \xi} &=& - \frac{i k''}{2}
\frac{\partial^{2} \mathcal{H}} {\partial t^2} + \frac{i}{2k_0}
\nabla_\perp^{2}\mathcal{H} + \frac{i 3 k_0}{2}
\left(\frac{\chi^{(3)}_e}{\epsilon_r}|\mathcal{E}|^{2} +
\frac{\chi^{(3)}_m}{\mu_r}|\mathcal{H}|^{2}\right)\mathcal{H}
\label{hampl}
\end{eqnarray}
where, in the left hand side, the transformation $\xi=z$, $\tau =
t-k'z$, was used; the transverse Laplacian
$\nabla_\perp=(\frac{\partial^2}{\partial
x^2},\frac{\partial^2}{\partial y^2})$ is defined in the plane
perpendicular to the $z$ axis.  The relative permittivity and the
relative permeability, $\epsilon_r$ and $\mu_r$, are evaluated in
$\omega_0$. The non-linear electric and magnetic susceptibilities,
$\chi^{(3)}_e$ and $\chi^{(3)}_m$, are the Fourier transforms
evaluated in $(\omega_0,\omega_0,-\omega_0)$. Dissipation (linear or
non-linear) is neglected, so that $\chi^{(1)}_e$, $\chi^{(1)}_m$,
$\chi^{(3)}_e$ and $\chi^{(3)}_m$ are real quantities. This is an
often met approximation in conventional optics, but dissipation could
play a relevant role in metamaterials.  In this section we will
derive the equations for a material without dissipation, and in
Sect.\ \ref{dissipation} we will analyze how these equations are
modified when dissipation is taken into account.

The nonlinear electric susceptibility is usually written as
$\chi^{(3)}_e = \alpha/E_c^2$, where $\alpha=\pm 1$ stands for a
focusing or defocusing nonlinearity and $E_c$ is a characteristic
electric field. Eqs.\ (\ref{eampl}) and (\ref{hampl}) are an
extension of the analysis performed in \cite{kockaert,kockaert2} to
include the magnetic non-linearity, and are similar to the ones
derived in \cite{lazarides}.

For an NRM, $k_0$ is negative, but this does not modify the sign of
the non-linear terms in Eqs.\ (\ref{eampl}) and (\ref{hampl}), since
$k_0/\epsilon_r$ and $k_0/\mu_r$ are always positive.  The difference
appears in the diffraction term, that becomes negative for an NRM.

\begin{figure}
\begin{center}
\includegraphics{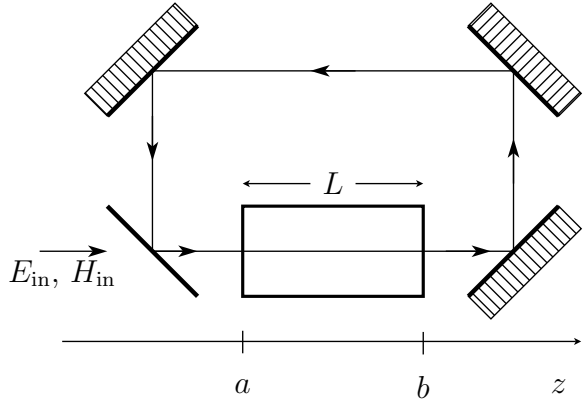}
\end{center}
\caption{Ring cavity with the nonlinear optic material of size $L$.}
\label{cavity}
\end{figure}

Fig.\ \ref{cavity} shows the geometry of the ring cavity.  At the
entrance mirror, the input field has amplitudes $E_\mathrm{in}$ and
$H_\mathrm{in}$, with $H_\mathrm{in} = E_\mathrm{in}
\sqrt{\epsilon_0/\mu_0}$. The non-linear material has length $L$, the
transmission coefficient in the right end of the material for the
electric field amplitude is $t_+=1+\zeta$, and for the magnetic field
amplitude is $t_-=1-\zeta$, where $\zeta
=(\eta-\eta_0)/(\eta+\eta_0)$ with $\eta_0=\sqrt{\epsilon_0/\mu_0}$
and $\eta=\sqrt{\epsilon/\mu}$. The transmission and reflection
coefficients in the input mirror are $t_i$ and $r_i$. The time of a
round trip is $T_r$, the phase accumulated by the wave in a round
trip is $\phi$, and the detuning between the pump field and the
cavity mode is $\theta = \phi\; \mathrm{mod}\; 2\pi$, with $\theta
\ll 1$ since we consider that the cavity is close to resonance.  It
is convenient to define the quantity $\rho = |r_i|(1-\zeta^2)$, the
length $l = \sqrt{L \rho/[2|k_0|(1-\rho)]}$, and the characteristic
electric field $C = t_+ E_c \sqrt{2\epsilon_r(1-\rho)/(3L k_0
\rho)}$. Using the non-linear Schr\"{o}dinger equations (\ref{eampl}) and
(\ref{hampl}) for the fields inside the cavity, and after the
following change of variables:
\begin{eqnarray}
A_1 = \frac{\mathcal{E}}{C}, \quad A_2 = \frac{\mathcal{H}}{C}
\sqrt{\frac{\mu_0}{\epsilon_0}}, \quad
x' = x/l, && y'= y/l, \quad
t'= t \frac{(1-\rho)}{T_r},\label{changevar}
\end{eqnarray}
we arrive to two Lugiato-Lefever type equations (the details of the
derivation can be found in the appendix):
\begin{eqnarray}
\frac{\partial A_1}{\partial t} = A_\mathrm{in} - (1+i\Theta) A_1 + i
\beta \nabla^2_\perp A_1 + i(\alpha |A_1|^2 + \gamma |A_2|^2) A_1
\label{lla1}\\
\frac{\partial A_2}{\partial t} = A_\mathrm{in} - (1+i\Theta) A_2 + i
\beta \nabla^2_\perp A_2 + i(\alpha |A_1|^2 + \gamma |A_2|^2) A_2
\label{lla2}
\end{eqnarray}
where $\Theta = -\theta \rho/(1-\rho)$, $\beta=\pm 1$ is the sign of
the refraction index $n$, $\gamma = \frac{\epsilon_0
\epsilon_r^2}{\mu_0 \mu_r^2} \chi_m^{(3)}/|\chi_e^{(3)}|$, and
$A_\mathrm{in} = \frac{t_i}{(1-\rho)} E_\mathrm{in}/C$.  The primes
have been omitted to simplify the notation.

Let us analyze the difference of the fields: $R=A_1 - A_2$. It is
possible to prove that $R$ decays exponentially to zero. The equation
for $R$ is
\begin{equation}
\frac{\partial R}{\partial t} = - (1+i\Theta) R + i \beta
\nabla^2_\perp R + i(\alpha |A_1|^2 + \gamma |A_2|^2) R,
\end{equation}
from which we get
\begin{equation}
\frac{\partial |R|^2}{\partial t} = -2|R|^2 +i\beta \bar{R}
\nabla^2_\perp R - i\beta R \nabla^2_\perp \bar{R}, \label{absr}
\end{equation}
where $\bar{R}$ is the complex conjugate of $R$.  We use
the expression of $R$ in terms of its Fourier transform,
$R(x,y,t) = \frac{1}{(2\pi)^2} \int dk_x\,dk_y R_{k_x,k_y}
e^{i(k_x x + k_y y)}$. The last two terms in Eq.\
(\ref{absr}) become
\begin{eqnarray}
\lefteqn{\bar{R} \nabla^2_\perp R - R \nabla^2_\perp \bar{R} =}
\nonumber \\
& & \frac{1}{(2\pi)^4} \int dk_x\,dk_y\,dq_x\,dq_y\, R_{k_x,k_y}
\bar{R}_{q_x,q_y} e^{i(k_x-q_x)x}\, e^{i(k_y-q_y)y} (q_x^2 + q_y^2 -
k_x^2 - k_y^2).\label{last2}
\end{eqnarray}
Let us consider the transverse average of Eq.\ (\ref{last2}), defined
as the integral over the plane $x$-$y$,
\begin{eqnarray}
\lefteqn{\langle \bar{R} \nabla^2_\perp R - R \nabla^2_\perp \bar{R}
\rangle =} \nonumber \\
& & \frac{1}{(2\pi)^4} \int dk_x\,dk_y\,dq_x\,dq_y\, R_{k_x,k_y}
\bar{R}_{q_x,q_y} (q_x^2 + q_y^2 - k_x^2 - k_y^2) \int dx\,dy\,
e^{i(k_x-q_x)x}\, e^{i(k_y-q_y)y} \nonumber \\
&=& \frac{1}{(2\pi)^2} \int dk_x\,dk_y\,dq_x\,dq_y\, R_{k_x,k_y}
\bar{R}_{q_x,q_y} (q_x^2 + q_y^2 - k_x^2 - k_y^2)\, \delta(k_x-q_x)\,
\delta(k_y-q_y) \nonumber \\
&=& 0.
\end{eqnarray}
Therefore, from Eq.\ (\ref{absr}) we get
\begin{equation}
\frac{\partial \langle |R|^2 \rangle}{\partial t} = -2\langle |R|^2
\rangle,
\end{equation}
that means that the transverse average $\langle |R|^2 \rangle$ decays
to zero exponentially with a characteristic time 1/2 (or
$T_r/[2(1-\rho)]$ for the previous time scale).  If, after a
transient, we have $\langle |R|^2 \rangle = 0$, then, since $|R|^2\ge
0$ in every point of the transverse plane, we have that $R=0$ in
every point. Then, after a time of order 1/2, we have that $A_1 =
A_2$. In this situation, the description of Eqs.\ (\ref{lla1}) and
(\ref{lla2}) is further simplified to
\begin{equation}
\frac{\partial A}{\partial t} = A_\mathrm{in} - (1+i\Theta) A + i
\beta \nabla^2_\perp A + i \gamma' |A|^2 A, \label{lla}
\end{equation}
where $A=A_1=A_2$ and $\gamma'=\alpha+\gamma$.  Eq.\ (\ref{lla}) has
the same form than the Lugiato Lefever equation, but there are some
differences. In the original version, that considers only an electric
non-linearity, it was shown \cite{lugiato} that the sign of the
detuning $\Theta$ must be equal to the sign of the non-linear
coefficient, $\gamma'$, but this is not necessarily the case in Eq.\
(\ref{lla}). In addition, in (\ref{lla}) the sign of the diffraction
term, $\beta$, can be negative or positive depending on the material
being an NRM or not, and the magnetic non-linearity is included in
the coefficient $\gamma'$.

In the particular case in which there is only an electric
non-linearity, i.e. $\gamma=0$, and using the result of
\cite{lugiato} that says that the sign of $\Theta$ is equal to
$\alpha$, that is equal to 1 ($-1$) for a focusing (defocusing)
electric non-linearity, we get,
\begin{equation}
\frac{\partial A}{\partial t} = A_\mathrm{in} - (1+i\alpha |\Theta|)
A + i \beta \nabla^2_\perp A + i \alpha |A|^2 A. \label{llelec}
\end{equation}
Eq.\ (\ref{llelec}) presents an interesting symmetry.  Taking its
complex conjugate, and assuming that $A_\mathrm{in}$ is real, it can
be seen that a focusing non-linearity and a positive refraction index
material ($\alpha=1$ and $\beta=1$) is equivalent to the defocusing
and NRM case ($\alpha=-1$ and $\beta=-1$).  Also the case $(\alpha=1,
\beta=-1)$ is equivalent to $(\alpha=-1, \beta=1)$.  As we will see
below, the equivalences are more involved when the magnetic
non-linearity is considered, since the signs of the three
coefficients ($\Theta$, $\beta$ and $\gamma'$) are in general not
related among them.

\section{Dissipation}
\label{dissipation}

 Although it is common to neglect dissipation in
conventional optics, this is not in general an appropriate
approximation for a metamaterial.  In this section we analyze how the
previous equations are modified when dissipation is taken into
account. In this case, the electric permittivity and magnetic
permeability are complex quantities. The wave-number that was defined
as $k(\omega) =
\frac{\omega}{c}\sqrt{\epsilon_r(\omega)\mu_r(\omega)}$ will also
have an imaginary part. At the frequency $\omega_0$ we have
\begin{equation}
k(\omega_0) = k_0 + i\, k_I
\end{equation}
where $k_0$ is the wave-number of the plane wave modulated by the
slowly varying amplitude (\ref{ansatz}), and $k_I$ is the imaginary
part.

We will consider that the attenuation distance is much larger than
the wave-length.  This means that $k_I \ll k_0$.  In the Gigahertz
range, it is possible to build an NRM with small (and even
negligible) imaginary parts of $\epsilon_r$ and $\mu_r$  as was shown
in, for example, \cite{liu,wang}.

It is assumed that $k_I$ is of second order in the small parameter
used in the multiple-scales perturbation technique, see \cite[p.
101]{moloney}. The result is an additional term in the non-linear
Schr\"{o}dinger equations for the envelopes of the electric and magnetic
fields
\begin{eqnarray}
\frac{\partial \mathcal{E}}{\partial \xi} &=& - \frac{i k''}{2}
\frac{\partial^{2} \mathcal{E}}{\partial t^2} + \frac{i}{2k_0}
\nabla_\perp^{2}\mathcal{E} + \frac{i 3 k_0}{2}
\left(\frac{{\chi^{(3)}_e}}{\epsilon_r}|\mathcal{E}|^{2} +
\frac{{\chi^{(3)}_m}}{\mu_r}|\mathcal{H}|^{2}\right)\mathcal{E} - k_I
\mathcal{E}
\label{eampld}\\
\frac{\partial \mathcal{H}}{\partial \xi} &=& - \frac{i k''}{2}
\frac{\partial^{2} \mathcal{H}} {\partial t^2} + \frac{i}{2k_0}
\nabla_\perp^{2}\mathcal{H} + \frac{i 3 k_0}{2}
\left(\frac{{\chi^{(3)}_e}}{\epsilon_r}|\mathcal{E}|^{2} +
\frac{{\chi^{(3)}_m}}{\mu_r}|\mathcal{H}|^{2}\right)\mathcal{H} - k_I
\mathcal{H}. \label{hampld}
\end{eqnarray}
The additional terms ($- k_I \mathcal{E}$ and $- k_I \mathcal{H}$)
represent an exponential decay of the amplitudes due to dissipation.
In Eqs.\ (\ref{eampld}) and (\ref{hampld}), the non-linear
coefficients $\chi^{(3)}_e/\epsilon_r$ and $\chi^{(3)}_m/\mu_r$ can
be taken as real quantities since the contribution of the imaginary
parts correspond to terms of higher order in the expansion of the
small parameter.

Following the steps indicated in the appendix, we arrive to a couple
of Lugiato-Lefever equations that have the same form that Eqs.\
(\ref{lla1}) and (\ref{lla2}), where now the change of variables is
given by
\begin{eqnarray}
A_1 = \frac{\mathcal{E}}{C'}, \quad A_2 = \frac{\mathcal{H}}{C'}
\sqrt{\frac{\mu_0}{\epsilon_0}}, \quad
x' = x/l', && y'= y/l', \quad
t'= t \frac{(1-\rho+\rho L k_I)}{T_r},  \label{changevar2}
\end{eqnarray}
with $C' = t_+ E_c \sqrt{2\epsilon_r(1-\rho+\rho L k_I)/(3L k_0
\rho)}$ and $l' = \sqrt{L \rho/[2|k_0|(1-\rho+\rho L k_I)]}$.  The
scaled detuning is now defined as $\Theta = -\theta \rho/(1-\rho+\rho
L k_I)$, and the input field is $A_\mathrm{in} =
\frac{t_i}{(1-\rho+\rho L k_I)} E_\mathrm{in}/C$.  The definitions of
the rest of the coefficients that appear in Eqs.\ (\ref{lla1}) and
(\ref{lla2}) remain the same.

Therefore, the case of small dissipation can still be correctly
described by Eqs.\ (\ref{lla1}) and (\ref{lla2}).  It is not
difficult to see that the demonstration of the previous section that
the amplitudes $A_1$ and $A_2$ are equal after a transient still
holds for this dissipative case. Consequently, the simplified
description of Eq.\ (\ref{lla}) also holds.

The situation is different if the attenuation distance is of order of
the wave-length, i.e., $k_I \sim k_0$, that corresponds to a factor
of merit of order one. This is generally the case for optical
frequencies. Now, the assumption that light can be represented by a
plane wave modulated by a slowly varying amplitude, Eq.\
(\ref{ansatz}), is no longer valid, because the decay of the fields
due to dissipation takes place in a fast space scale, and this decay
can not be described by the envelopes $\mathcal{E}$ and
$\mathcal{H}$.  The situation when the length of the material,
$L$, is much greater than the wave-length,  is not
interesting since no field will be detected at the output.  There are
many proposals to reduce losses in metamaterials, some of them are
\cite{dolling,shalaev,popov,tassin,schuller,zhu}.

\section{Stability of homogeneous stationary solutions}
\label{staban}

The homogeneous stationary solutions, $A_0$, of Eq.\ (\ref{lla}) are
obtained by solving
\begin{equation}
A_\mathrm{in}= (1+i\Theta-i \gamma' I_0) A_0,
\end{equation}
where $I_0=|A_0|^2$.  For the intensities we have
\begin{equation}
I_\mathrm{in} = (1+\Theta^2)I_0 - 2\Theta \gamma'I_0^2 + \gamma'^2
I_0^3, \label{intensity}
\end{equation}
where $I_\mathrm{in} = |A_\mathrm{in}|^2$ is the intensity of the
input field.  It is well known that Eq.\ (\ref{intensity}) presents
bistability for $|\Theta| > \sqrt{3}$, i.e., there is a range of
values of $I_\mathrm{in}$ for which there are two stable solutions of
$I_0$.  Note that, since $\Theta = -\theta \rho/(1-\rho)$, and
$\theta \ll 1$, bistability can only be attained when $\rho =
|r_i|(1-\zeta^2) \simeq 1$, i.e., when the entrance mirror has a high
reflectivity and the material has a high transmissivity.  Another
condition to have bistability is $\Theta \gamma' > 0$, this condition
is automatically fulfilled when there is only an electric
non-linearity, i.e., when $\gamma=0$ and $\Theta=\alpha |\Theta|$.

In terms of the solutions for the intensity $I_0$ obtained from
(\ref{intensity}), the homogeneous solution is
\begin{equation}
A_0 = (1 - i\Theta + i \gamma'I_0)\,I_0/A_\mathrm{in},
\end{equation}
where we have assumed, without loss of generality, that
$A_\mathrm{in}$ is real.

\begin{figure}
\begin{center}
\includegraphics[width=12cm]{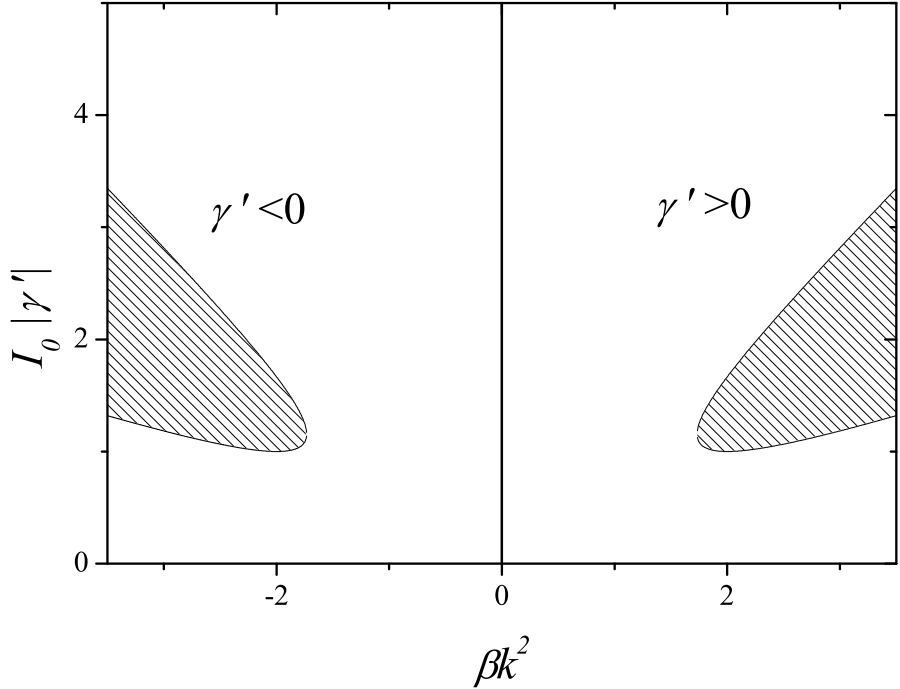}
\end{center}
\caption{Marginal stability curves, $I_0\,|\gamma'|$ versus $\beta
k^2$, for $\Theta=0$. In the dashed region the homogeneous solution
becomes unstable.} \label{marginal}
\end{figure}

A linear stability analysis of $A_0$ gives the following eigenvalues,
\begin{equation}
\lambda_\pm = -1\pm\sqrt{[3\gamma'I_0-\beta k^2-\Theta][\beta k^2 +
\Theta - \gamma'I_0]}, \label{eigen}
\end{equation}
where $k$ is the wave-number of a small perturbation. From
(\ref{eigen}) we can draw the marginal stability curves of Fig.\
\ref{marginal}.  The same figure applies to both cases: negative and
positive refraction index, that correspond to the negative and
positive parts, respectively, of the horizontal axis. In the regions
enclosed by the curves, the homogeneous solution $A_0$ becomes
unstable.  Both regions do not appear simultaneously: one corresponds
to $\gamma' >0$ and the other to $\gamma' < 0$.  A value of $\Theta
=0$ was used in the figure, but a different value of $\Theta$ only
represents a horizontal shift of the curves.  The homogeneous
solution becomes unstable when the intensity is increased from zero
an reaches the value $I_0\,|\gamma'| = 1$. Over the instability
threshold, and close to it, it is known that a hexagonal pattern
appears \cite{scroggie,tlidi}, when the critical wave number is
different from 0 (this happens when there is not bistability or, more
strictly, when $\beta\Theta < 2$).

Eq.\ (\ref{lla}) and its complex conjugate represent the same
physical situation.  The only difference present in the complex
conjugate equation is the sign in front of the three real parameters
$\Theta$, $\beta$ and $\gamma'$.  Using this equivalence we can
identify three typical situations: (a) $\beta \gamma'>0$ and any
value of $\Theta$; (b) $\beta \gamma'<0$ and $\Theta \gamma'<0$; and
(c) $\beta \gamma'<0$ and $\Theta \gamma'>0$.

In case (a) $\beta$ and $\gamma'$ have the same sign.  We can see in
Fig.\ \ref{marginal} that, for $\beta=-1$ or 1, the marginal
stability curve is present, so that, as the input field intensity is
increased, the homogeneous solution becomes always unstable.  The
critical values of field intensity and wave-number are given by,
\begin{eqnarray*}
I_{0,c}|\gamma'| =1 \ \ \mathrm{and}\ \ k_c^2=2-|\Theta| &&\
\mathrm{for}\ \beta \Theta < 2 \\
I_{0,c}|\gamma'| = 2|\Theta|/3 - \sqrt{\Theta^2-3}/3 \ \
\mathrm{and}\ \ k_c=0 &&\ \mathrm{for}\ \beta \Theta > 2
\end{eqnarray*}

In case (b), the homogeneous solution is always stable.  Bistability
is not possible in this case since $\Theta \gamma'< 0$.

In case (c), the homogeneous solution becomes unstable only if there
is bistability, i.e., when $|\Theta|>\sqrt{3}$.  The critical values
are,
\begin{eqnarray*}
I_{0,c}|\gamma'| = 2|\Theta|/3 - \sqrt{\Theta^2-3}/3 \ \
\mathrm{and}\ \
k_c=0 &&\ \mathrm{for}\ \sqrt{3}<|\Theta|<2 \\
I_{0,c}|\gamma'| =1 \ \ \mathrm{and}\ \ k_c^2=|\Theta|-2 &&\
\mathrm{for}\ |\Theta|>2
\end{eqnarray*}

\begin{figure}[ht!]
\begin{center}
\includegraphics[width=12cm]{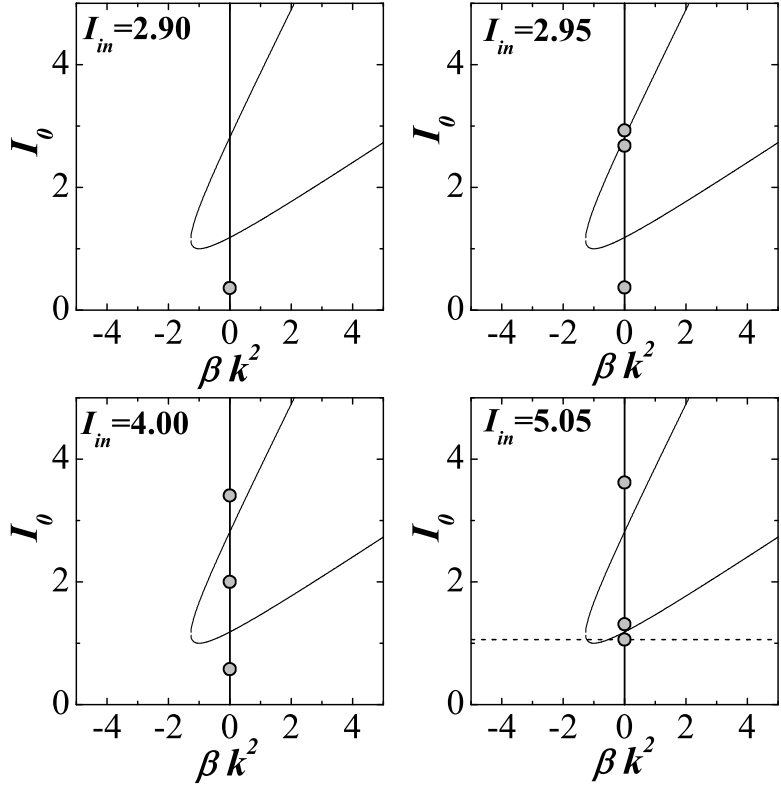}
\end{center}
\caption{Marginal stability curve, $I_0$ versus $\beta k^2$, for the
bistability case with $\Theta=3$, and $\gamma'=1$. The sequence
corresponds to increasing values of $I_\mathrm{in}$, from left to
right and from top to bottom: $I_\mathrm{in} = 2.9$, 2.95, 4 and
5.05. The small circles on the vertical axis show the positions of
the homogeneous solutions. The negative horizontal axis corresponds
to case (c) and the positive one to case (a).} \label{marseq}
\end{figure}

If $\Theta > \sqrt{3}$, and $\gamma'>0$, the vertical axis in Fig.\
\ref{marginal} is shifted to the right and crosses the marginal
stability curve in two points that give a range of values of $I_0$
for which the solution $A_0$ becomes unstable under homogeneous
perturbations. This corresponds to the unstable branch when
bistability is present. Fig.\ \ref{marseq} shows a sequence of plots
with the marginal stability curve with the values of the homogeneous
solutions indicated on the vertical axis for increasing values of the
input intensity $I_\mathrm{in}$ and for $\Theta=3$ and $\gamma'=1$.
As $I_\mathrm{in}$ increases from 2.9 to 2.95 a saddle node
bifurcation takes place in the upper branch of the marginal stability
curve (the bifurcation happens in the two dimensional phase space
composed by the real and imaginary parts of $A_0$). As
$I_\mathrm{in}$ is further increased, the value of the intensity of
the homogeneous solution of the unstable branch (the one that is
inside the instability region delimited by the curve) decreases,
until it merges, in another saddle node bifurcation, with the lower
homogeneous solution.

We have an example of case (c) in Fig.\ \ref{marseq} for $\beta=-1$,
where the lower homogeneous solution becomes unstable before it
merges with the homogeneous solution of the unstable branch (see
dotted line in bottom right plot of Fig.\ \ref{marseq}). The saddle
node bifurcation in the lower branch takes place at $I_0$
approximately equal to 1.18 for the parameters of Fig.\ \ref{marseq},
but a modulational instability with a wave-number different from zero
takes place at a lower value of $I_0$ ($I_0=1$).

\begin{figure}[ht!]
\begin{center}
\includegraphics[width=12cm]{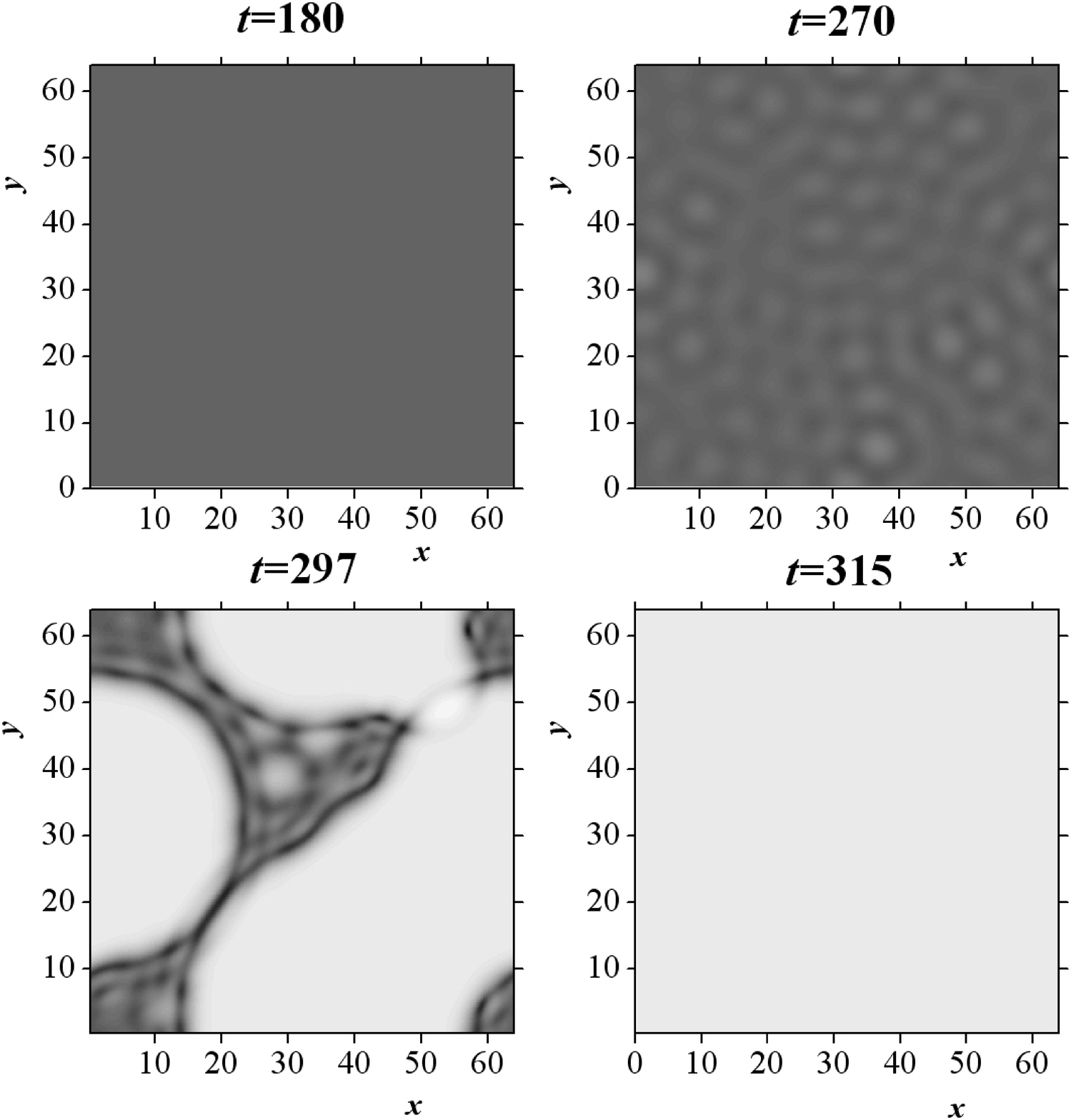}
\end{center}
\vspace{-0.5cm}
\caption{Four configurations of $|A|^2$ against position in
the transverse plane, obtained from integration of Eq.\
(\ref{lla}) for increasing values of time.  Parameters are
$\beta=-1$, $\Theta=3$ and $\gamma'=1$, corresponding to
case (c). The initial condition has intensity $I_0=1.03$
($I_\mathrm{in}=5.02$), slightly above the critical value
($I_{0,c}=1$). The system evolves from the lower to the
upper homogeneous solution. The space and time integration
steps are $\Delta x = 0.25$ and $\Delta t = 0.003$.  The
grey scale is logarithmic, black corresponds to 0.4 and
white to 4.5.} \label{numh-1}
\end{figure}

Fig.\ \ref{numh-1} shows four snapshots of numerical integration
results for this case. The initial condition is the lower homogeneous
solution slightly above the instability threshold ($A_0 = 0.457 - i
0.908$, i.e. $I_0 = 1.03$), plus noise of amplitude 0.005. Eq.\
(\ref{lla}) was integrated in a square domain of 256$\times$256
points with periodic boundary conditions, using a Fourier series
representation and a 4th order Runge-Kutta temporal scheme for the
non-linear terms. At time $t=180$ the field still appears
homogeneous. At $t=270$ a pattern characterized by the critical wave
number appears ($k_c=1$). At $t=297$ we can see some spatial domains
where the field takes the value of the upper homogeneous solution.
These domains grow until the whole system becomes homogeneous.
Therefore, the numerical results indicate that, as the input
intensity is increased, the lower homogeneous solution becomes
unstable, but this instability does not give rise to a stable
non-homogeneous pattern. Instead, the system evolves to the upper
homogeneous solution, that is always stable for $\beta=-1$.

\begin{figure}
\begin{center}
\includegraphics[width=12cm]{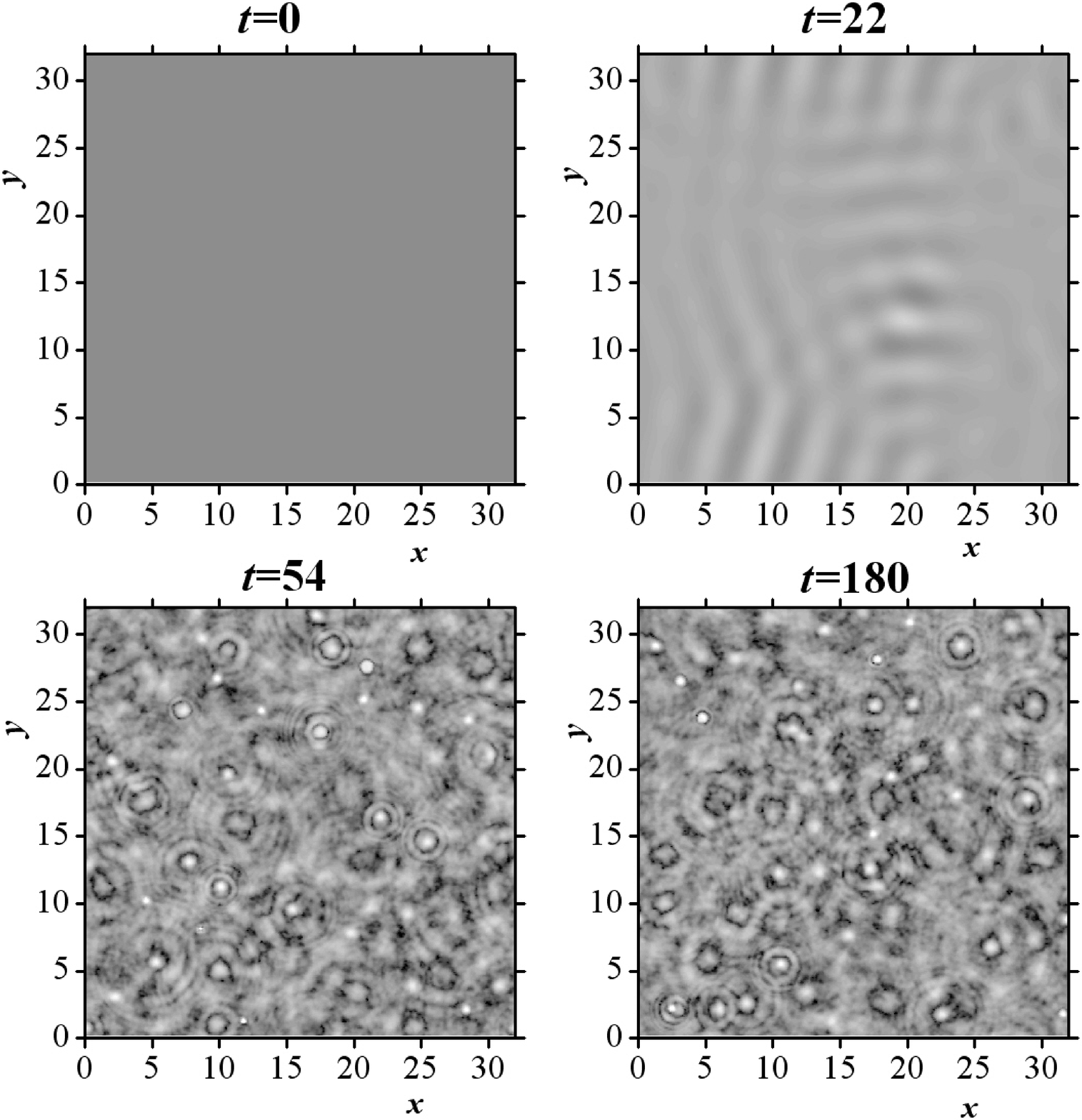}
\end{center}
\vspace{-0.5cm}
\caption{Intensity $|A|^2$ for increasing values of time calculated
from Eq.\ (\ref{lla}).  Parameters are $\beta=1$, $\Theta=3$ and
$\gamma'=1$, corresponding to case (a).  The homogeneous initial
condition has intensity equal to 1.19 plus noise of amplitude 0.005,
and the input field has $I_\mathrm{in}=5.11$. The final state
corresponds to spatiotemporal chaos that is reached independently of
the value of the initial homogeneous state.  Space and time
integration steps: $\Delta x = 0.125$ and $\Delta t = 0.001$. The
gray scale is logarithmic with black equal to 0.1 and white equal to
60. We used a different scale for $t=22$: black=3, white=4, in order
to show the transient emerging pattern.} \label{numh1}
\end{figure}

Continuing with the analysis of Fig.\ \ref{marseq}, the situation is
different for $\beta=1$, that corresponds to case (a); in this case
the lower homogeneous solution is always stable until the saddle node
bifurcation takes place.  Therefore, as the input intensity is
increased, the lower homogeneous solution does not simply cross the
marginal stability curve and lose its stability, but, instead, it
does no longer exist as a stationary solution since the homogeneous
mode, $k=0$, becomes unstable. We performed numerical integration of
Eq.\ (\ref{lla}) for an input intensity slightly above the value
corresponding to the saddle node bifurcation.  For different values
of the homogeneous initial condition, the evolution is qualitatively
similar to the one shown in Fig.\ \ref{numh1}. Initially, only the
modes close to 0 are unstable, so that, for short times, the value of
the intensity increases keeping, approximately, the homogeneous shape
of the field. For larger times, the system evolves to a state of
optical turbulence or spatiotemporal chaos.

As mentioned before, in case (a) without bistability, an hexagonal
pattern appears close and above the instability threshold.  It was
shown in Ref.\ \cite{gomila} that, as the input intensity is
increased, there is a sequence of different spatiotemporal regimes:
oscillating hexagons, quasiperiodicity, temporal chaos and optical
turbulence. When there is bistability, the state of optical
turbulence is reached directly as a transition from the homogeneous
state as the input intensity is increased (see Fig.\ \ref{numh1}).

\begin{figure}[ht!]
\begin{center}
\includegraphics[width=10cm]{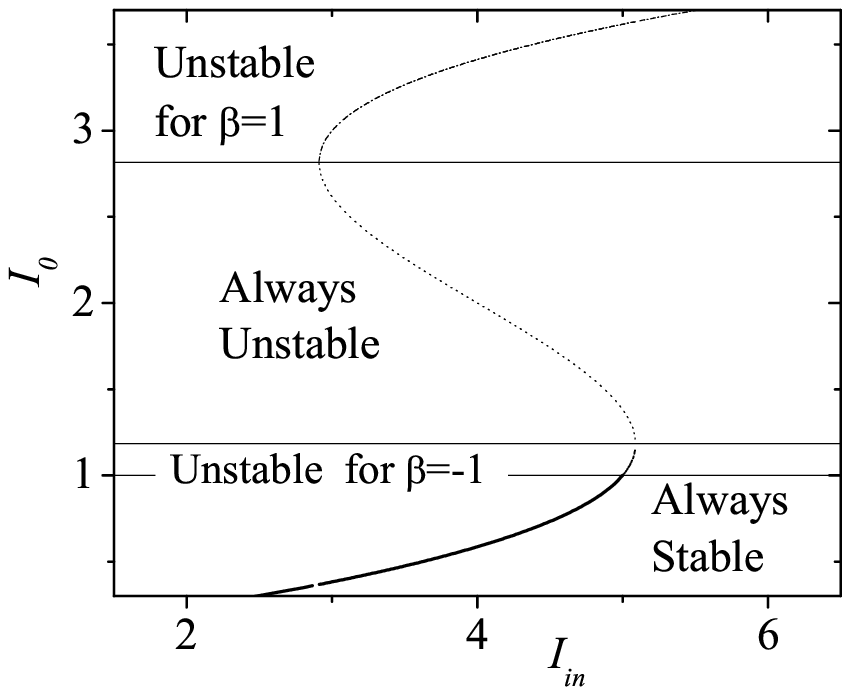}
\end{center}
\vspace{-0.5cm}
\caption{Intensity of the homogeneous solution $I_0$ against
intensity of the input field $I_\mathrm{in}$ in the bistability case,
with $\Theta=3$ and $\gamma'=1$. The continuous curve corresponds to
an always stable homogeneous solution.  In the rest of the curve, the
stability range depends on the sign of the refraction index, $\beta$.
The part with negative slope (dotted curve) is always unstable for
$\beta=1$ or $-1$.  The upper part (dash-dotted curve) is unstable
only for $\beta=1$ but it is stable for an NRM ($\beta=-1$).
Equivalently, for the region $1<I_0 < 1.18$ the homogeneous solution
is unstable for $\beta=-1$ but is stable for $\beta=1$.}
\label{homosol}
\end{figure}

The differences between the stability of homogeneous solutions for
positive ($\beta=1$) or negative ($\beta=-1$) refraction index
materials, when bistability is present, is clearly presented in Fig.\
\ref{homosol}. The figure shows the curve $I_0$ against
$I_\mathrm{in}$.  The shape of the curve is the same for $\beta=1$ or
$\beta=-1$, but the stability ranges are different.

The stability analysis is similar to the one presented in
\cite{kockaert}, where only an electric non-linearity was considered.
The inclusion of the magnetic non-linearity is explicitly represented
by factor $\gamma$ in (\ref{lla}).  In addition, a result that, to
our knowledge, was not previously reported, is the equivalence
between positive and negative refraction index materials when the
signs of the detuning $\Theta$ and the non-linear coefficient
$\gamma'$ are changed.  This symmetry is illustrated, for example, in
Fig.\ \ref{marginal}, for $\Theta=0$. In this case, the behavior for
$\beta=1$ and $\gamma'>0$ is equivalent to the behavior for
$\beta=-1$ and $\gamma'<0$.

\section{Conclusions}
\label{concl}

Starting from the Maxwell equations, with non-linear polarization and
magnetization, it is possible to obtain, using the multiple-scales
technique, two coupled non-linear Schr\"{o}dinger equations for the
electric and magnetic field amplitudes.  From these equations we
derived the evolution of the fields in a ring cavity with plane
mirrors containing a material with positive or negative refraction
index and with effective electric and magnetic non-linearities.  We
proved that the description can be reduced to only one equation that
has the same form of the Lugiato Lefever equation \cite{lugiato} for
a positive refraction index material with only electric
non-linearity.

An original contribution of the paper is the generalization of the
Lugiato Lefever equation to electric and magnetic non-linearities.
Let us note that, as was shown in \cite{zharov}, effective
macroscopic magnetic non-linearities can become relevant in composite
materials that are used to generate a negative refraction index. The
resulting equation has three real parameters: the sign of the
refraction index $\beta$, the detuning $\Theta$, and the non-linear
coefficient $\gamma'$.  In the original version of the Lugiato
Lefever equation, the sign of the detuning is equal to the sign of
the non-linear coefficient, corresponding to the self-focusing or
self-defocusing cases.  In the present version, both signs are
independent; in addition, the diffraction coefficient can be positive
or negative depending on $\beta$.  This generalization of the
parameters allows the equation to represent new physical situations.
Despite the fact that more free parameters necessarily makes the
analysis more complex, using a linear stability analysis we have
shown that any combination of the parameters must correspond to one
of only three typical cases.  In case (b), $\beta \gamma'<0$ and
$\Theta \gamma'<0$, the homogeneous solution is always stable. In
case (c), $\beta \gamma'<0$ and $\Theta \gamma'>0$, the homogeneous
solution can become unstable only when there is bistability, but, in
this case, numerical integration shows that the final state is the
homogeneous solution of the upper branch. Only in case (a), $\beta
\gamma'>0$, the destabilization of the homogeneous solution, as the
input field intensity is increased, gives rise to a non-homogeneous
state.  In this last case, if there is not bistability it is known
that, close to the instability threshold, the asymptotic state is an
hexagonal pattern. If there is bistability, numerical results for
$\Theta=3$ show that there is a transition from the homogeneous state
to optical turbulence as the input field is increased.

In summary, the description, in the plane perpendicular to
propagation, of the evolution of the electromagnetic field in a
cavity with a non-linear material with positive or negative
refraction index has been reduced to a Lugiato Lefever equation with
three parameters: $\Theta$, $\gamma'$ and $\beta$. These quantities
are functions of the much larger set of parameters of the original
description (based on the Maxwell equations). One of the aims of the
work was the identification of relevant parameters since they allow a
better understanding of typical behaviours that the system can
develop.  This kind of analysis is much more difficult to perform
with the original description.

\subsection*{Acknowledgements}
We thank Carlos Martel for helpful discussions. This work was
partially supported by Consejo Nacional de Investigaciones
Cient\'{\i}ficas y T\'{e}cnicas (CONICET, Argentina), Agencia Nacional de
Promoci\'{o}n Cient\'{\i}fica y Tecnol\'{o}gica ANPCyT (PICT 2004, N 17-20075,
Argentina) and Universidad Polit\'{e}cnica de Madrid (Spain) under grant
AL09-P(I+D).

\section*{Appendix}

In this appendix we present the derivation of the cavity equations
(\ref{lla1}) and (\ref{lla2}) starting from the non linear
Schr\"{o}dinger equations (\ref{eampl}) and (\ref{hampl}).  It is
essentially an extension of the procedure described in
\cite{kockaert} to the case of electric \emph{and} magnetic
non-linearity.

Using Eqs.\ (\ref{eampl}) and (\ref{hampl}), for a small distance
$\delta \xi$, we can write
\begin{equation}
\mathcal{F}_\pm(\xi+\delta \xi)  = \mathcal{F}_\pm(\xi) + i \,\delta
\xi\, \hat{N} \mathcal{F}_\pm(\xi) \simeq e^{i\,\delta \xi\, \hat{N}}
\mathcal{F}_\pm(\xi) \label{ap23}
\end{equation}
where $\mathcal{F}_+ = \mathcal{E}$, $\mathcal{F}_- = \mathcal{H}$,
and $\hat{N}$ is the operator
\begin{equation}
\hat{N} = - \frac{k''}{2} \frac{\partial^2}{\partial t^2} +
\frac{1}{2 k_0} \nabla_\perp^{2} + \frac{3 k_0}{2}
\left(\frac{\chi^{(3)}_e}{\epsilon_r}|\mathcal{E}|^{2} +
\frac{\chi^{(3)}_m}{\mu_r}|\mathcal{H}|^{2}\right) \label{defN}
\end{equation}

Fig.\ \ref{cavity} shows the scheme of the cavity with the non linear
material of length $L$ whose left and right ends are at positions $a$
and $b$.  We call $a_+$ and $a_-$ the positions immediately to the
right and to the left of $a$ respectively, and similarly for $b_+$
and $b_-$.  Using the impedance of free space,
$\eta_0=\sqrt{\epsilon_0/\mu_0}$, and of the material,
$\eta=\sqrt{\epsilon/\mu}$, the transmission coefficient for the
electric field amplitude, when light goes from $b_-$ to $b_+$, is
\begin{equation}
t_+ = \frac{2\eta}{\eta + \eta_0} = 1+\zeta
\end{equation}
where $\zeta = (\eta-\eta_0)/(\eta+\eta_0)$ is a small quantity, so
that the transmission coefficient is close to 1. (Even when $\eta$ is
not close to $\eta_0$, a transmission coefficient close to 1 can be
achieved by filling the cavity with a substance with an impedance
close to $\eta$.)  The transmission coefficient for the magnetic
field amplitude in $b$ is,
\begin{equation}
t_- = \frac{2\eta_0}{\eta + \eta_0} = 1-\zeta.
\end{equation}
Using also the transmission and reflection coefficients in the input
mirror, $t_i$ and $r_i$, we obtain the following relations
\begin{eqnarray}
\mathcal{F}_\pm(b_+) &=& t_\pm \mathcal{F}_\pm(b_-) \nonumber \\
\mathcal{F}_\pm(a_+) &=& t_\mp \mathcal{F}_\pm(a_-) \nonumber \\
\mathcal{F}_\pm(a_-) &=& |r_i| e^{i\phi} \mathcal{F}_\pm(b_+) + t_i
F_{\mathrm{in}\pm}\label{ap24}
\end{eqnarray}
where $F_{\mathrm{in}+}=E_{\mathrm{in}}$,
$F_{\mathrm{in}-}=H_{\mathrm{in}}$, and $\phi$, the phase accumulated
in a round trip, includes the phase change of $4\pi$ due to
reflection in the four mirrors.  We will consider that the system is
close to resonance so that the detuning is $\theta \ll 1$, with
$\theta = \phi \ \mathrm{mod}\ 2\pi$.  Let $\mathcal{F}_{*\pm}$ be
the value of $\mathcal{F}_\pm$ after one round trip in the cavity,
that takes a time $T_r$. Applying the relations (\ref{ap23}) and
(\ref{ap24}) we find (subindices $\pm$ are removed for simplicity)
\begin{eqnarray}
\mathcal{F}_*=  t_i F_{in} + |r_i| (1-\zeta^2) e^{i\theta} e^{i L
\hat{N}} \mathcal{F} \simeq t_i F_{in} + |r_i|(1-\zeta^2) (1 +
i\theta + i L \hat{N}) \mathcal{F}.\label{ap25}
\end{eqnarray}
According to (\ref{defN}), the operator $\hat{N}$ depends on the
fields inside the material, but, in Eq.\ (\ref{ap25}), $\mathcal{F}$
corresponds to the fields outside the material, so $\hat{N}$ should
be evaluated in $t_\mp \mathcal{F}_\pm$. Writing the time derivative
of $\mathcal{F}$ as $(\mathcal{F}_*- \mathcal{F})/T_r$, and defining
$\rho = |r_i|(1-\zeta^2)$, we obtain,
\begin{equation}\label{ap26}
 T_r \frac{\partial \mathcal{F}}{\partial t} = t_i F_{in} +
 (\rho-1)\mathcal{F}+ i \rho \theta \mathcal{F}+
 i \rho L \, \hat{N}\mathcal{F}.
\end{equation}
After replacing the operator $\hat{N}$ by its definition
(\ref{defN}), and using the change of variables defined by Eqs.\
(\ref{changevar}), we arrive to Eqs.\ (\ref{lla1}) and (\ref{lla2})
for the amplitudes of the electric and magnetic fields in the
cavity.  It can be shown that, in the cavity equations, the
dispersion term proportional to $k''$ in (\ref{defN}) can be
neglected.

\end{document}